\documentclass{JINST}

\usepackage{amssymb}

\usepackage{graphics}
\usepackage{graphicx}

\title{Development of one-coordinate gaseous detector for wide angle diffraction studies.}

\author{Aulchenko V.M.$^a$, Papushev P.A.$^a$, Sharafutdinov M.R.$^b$,
Shekhtman L.I.$^a$\thanks{Corresponding author}, Titov
V.M.$^a$, Tolochko B.P.$^b$, Zhulanov V.V.$^a$ \\
\llap{$^a$}Budker Institute of
Nuclear Physics\\ 11 Lavrentiev Avenue, Novosibirsk 630090\\
Russia. Fax: 7(383)3307163,\\ e-mail: L.I.Shekhtman@inp.nsk.su\\
\llap{$^b$}Institute of Solid-State Chemistry and
Mechano-Chemistry\\
630090 Novosibirsk, Russian Federation}

\abstract {A one-coordinate gaseous detector of soft X-ray photons
for wide-angle X-ray scattering (WAXS) studies is being developed.
The detector operates in counting mode and is based on multi-stage
Gas Electron Multiplier(GEM). Full size detector is assembled and
has 67 degrees aperture with 350mm distance to the source, readout
multi-strip structure with 2048 strips at 0.2mm pitch and is
partially equipped with readout electronics in the central part.
Main parameters of the detector have been measured with 8keV X-ray
beam at VEPP-3 synchrotron ring. Spatial resolution of 470 $\mu$m
(FWHM) has been demonstrated that will allow to distinguish
diffraction spots at 0.1 degrees. }


\begin{document}

\section{Introduction.}
\label{OD4-intro}

Any detector for the studies of X-ray diffraction to large angles
has to present essentially curved geometry with conversion,
readout and possibly amplifying structure that surrounds the
scattering source. The detector OD3 with angular aperture up to
$30^{o}$ for powder diffraction experiments has been already
designed and constructed in Budker Institute of Nuclear
Physics~\cite{OD3-descr}. This detector has been built on the
basis of wire chamber. However further development of a detector
with larger angular aperture and considerably curved cylindrical
geometry is impossible on such basis. Thus in order to cover the
wider range of diffraction angles  the Gas Electron
Multiplier~(GEM)~\cite{GEM-Sauli} was proposed to be used as the
multiplying element.

GEM is a thin plastic foil double clad with copper layers from
both sides and pierced with regular array of small diameter holes.
Regular GEM is produced of 50$\mu$m thick kapton foil and has
140$\mu$m holes pitch and 80$\mu$m holes diameter. Gas
amplification occurs in the GEM holes when high voltage is applied
between the two foil sides. GEMs can be cascaded and in a
triple-GEM cascade can provide stable gain up to and higher than
$\sim10^5$ in a regular gas mixures like $Ar-CO_2$(70-30)
~\cite{GEM-highgain}.

Flat and flexible amplifying structure of GEM allows to prepare
arc-shaped GEM cascade that can surround the scattering source in
a WAXS experiment. Such approach can solve the problem of large
angular aperture for a gaseous detector.

The first measurements with the small prototype and simulations
demonstrating feasibility of this approach were described
elsewhere (~\cite{OD4-paper1}, ~\cite{OD4-SNIC06}). This paper
presents the first results obtained with full-size detector.

\section{Detector design and experimental set-up.}

\begin{figure}[htb]
\centering
\includegraphics[width=0.6\textwidth]{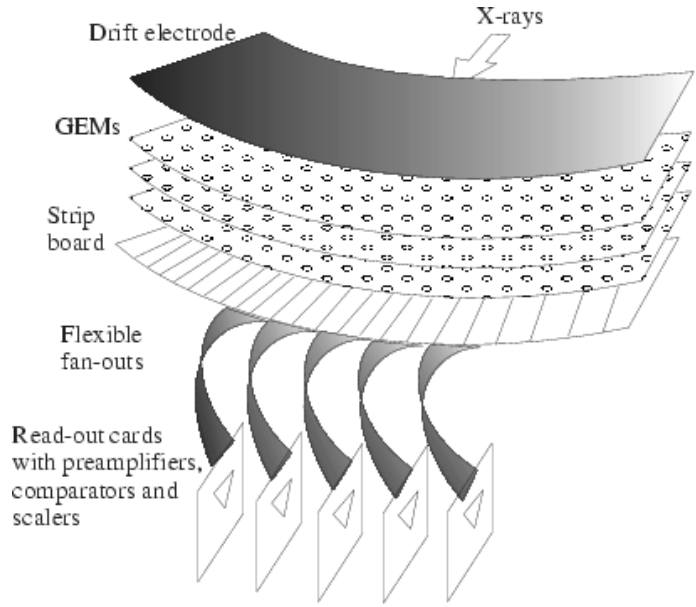}
\caption{Schematic view of OD4 design.} \label{fig:OD4-design}
\end{figure}

The detector for WAXS studies based on cascaded GEM (OD4) is shown
schematically in Fig.~\ref{fig:OD4-design}. X-rays from the
scattering source get into the gas box through the Be window and
are absorbed in 5.5mm thick drift gap between the drift cathode
and the top GEM. The triple-GEM stack with GEM to GEM distance of
1.5mm is attached on top of the multi-strip PCB at a distance of
2.5mm. Drift cathode, GEMs and PCB have arc shape with the center
at the scattering source. Strips of the PCB are positioned along
radii of the circle with the center at the source. The PCB
contains 2048 strips with the pitch of 0.2mm at the entrance side.

The detector is intended to work with soft X-rays in the range of
5keV to 15kev and is filled with $Ar-CO_2$(3:1) mixture at
atmospheric pressure.

The OD4 electronics is implemented on the basis of
preamplifier-shaper chip IC31A~\cite{BNL-TEC-PS} developed for the
electronics of PHENIX detector at RHIC (BNL, USA). Each strip of
the PCB is connected to an input of the preamplifier-shaper
through the flexible cable. The outputs of preamplifiers are
connected to the comparators with single and adjustable threshold
and logical pulses after the comparators are counted by scalers.

\begin{figure}[htb]
\centering
\includegraphics[width=0.7\textwidth]{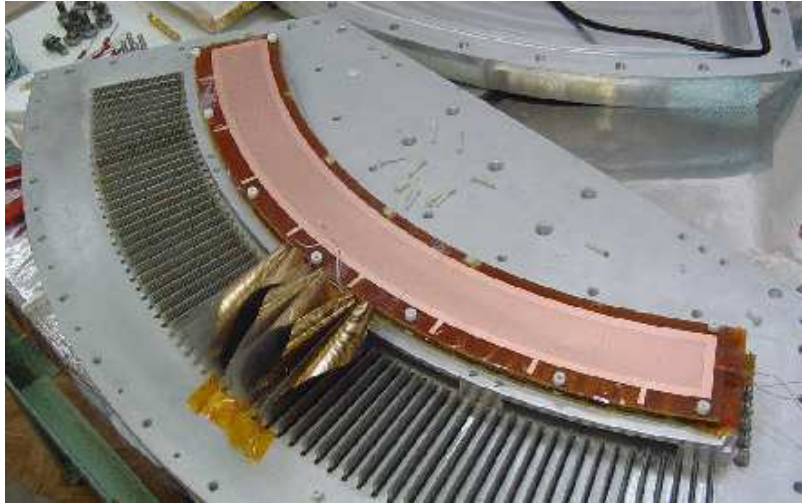}
\caption{Photo of the detector during assembly. Flexible kapton
cables are connecting 160 strips to the feed-throughs in the
central part of the detector. Triple-GEM cascade is installed on
top of the PCB. \newline \newline \newline \newline \newline
\newline \newline \newline } \label{fig:OD4-view1}
\end{figure}

\begin{figure}[htb]
\centering
\includegraphics[width=0.7\textwidth]{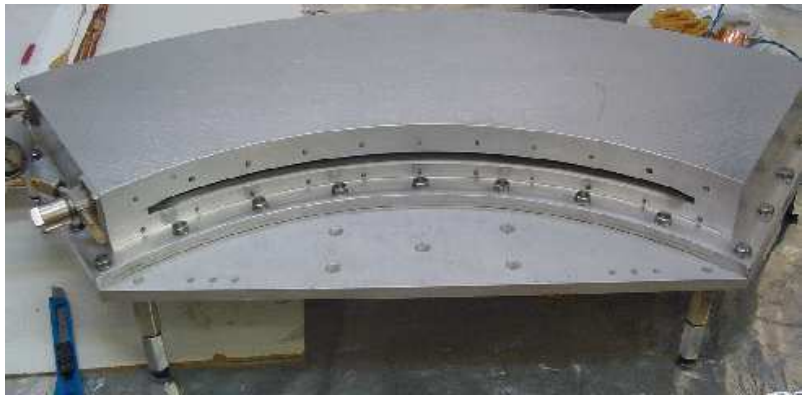}
\caption{View of the assembled detector. } \label{fig:OD4-view2}
\end{figure}

At present the final electronics is not yet ready and only 32
strips in the center of the detector have been equipped with
preamplifier-shapers and comparators. 64 strips from both sides of
the equipped area have been connected to ground to ensure uniform
field in the central region. The detector during assembling is
shown in Fig.~\ref{fig:OD4-view1} where triple-GEM cascade
installed on top of the PCB can be observed.
Fig.~\ref{fig:OD4-view2} demonstrates assembled detector.

The GEM electrodes and drift cathode have been powered through the
single-line resistive divider that was adjusted to minimize
transverse diffusion and have reasonable GEM transparency during
electrons drift from drift gap to the PCB. The adjustment of the
voltages across transfer gaps, induction gap and drift gap has
been performed for the conditions when the voltages across GEMs
provide total effective gain of the whole cascade around 10000.

For the measurements described in this paper the OD4 has been
installed at one of synchrotron radiation lines at VEPP-3 electron
ring. After monochromator the X-ray beam with 8.3 keV energy was
collimated by 20$\mu$m slit. Effective beam size at the entrance
window of the detector was 20$\mu$m in horizontal and $\sim$1mm in
vertical. The detector was positioned with 3 stands that allow
precise rotation around vertical axis, movement in horizontal and
vertical directions. With these stands the strips equipped with
the electronics could be aligned along the beam as the source was
not at the focus of the detector(350mm).

\section{Results and discussion}

\begin{figure}[htb]
\centering
\includegraphics[width=0.6\textwidth]{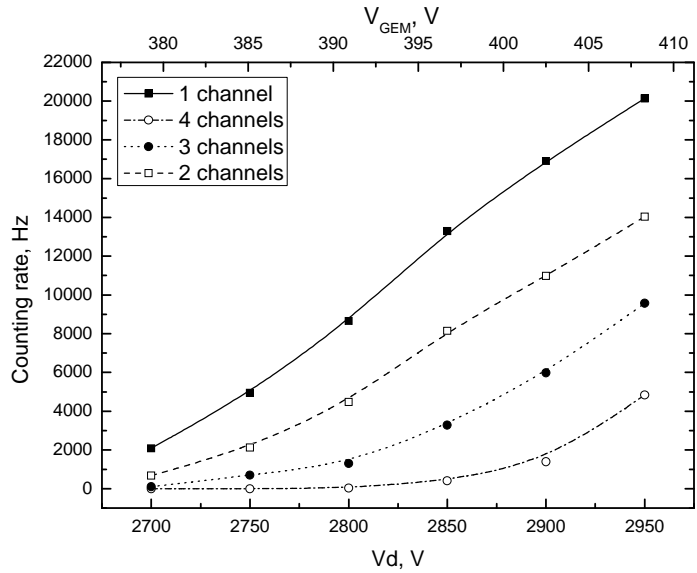}
\caption{Counting rate as a function of voltage at the resistive
divider while the detector is uniformly irradiated by 5.9keV
photons. The voltage across each GEM is shown at the top scale.
The rate of single counts, double-, triple- and quadruple-
coincidences is shown. } \label{fig:Fe55-count}
\end{figure}

When an X-ray photon is absorbed in the drift gap of the detector,
after charge transport and amplification the final charge cluster
occupies in average more than 1 strip. The charge distribution and
its effect on spatial resolution was discussed in details in our
previous paper(~\cite{OD4-paper1}) where the results of
simulations were compared to the measurements with the prototype
of the present detector. However the important outcome of the
charge distribution over several strips is that even if the
detector is irradiated with the constant flux of photons, the
counting rate will depend on the comparators threshold, gas gain
and the flux distribution in space. Fig.~\ref{fig:Fe55-count}
demonstrates the counting rate as a function of voltage at the
resistive divider in 1 arbitrary channel, coincidence between 2,3
and 4 neighboring channels while the detector has been uniformly
irradiated with $Fe^{55}$ 5.9keV photons. The comparators
threshold has been close to 150mV in this measurement that,
according to the electronic calibration (measured amplifier gain
was 14mV/fC), corresponded to the gas gain of $\sim$300 for 5.9keV
X-rays. Counting rate of a single channel is increasing in the
whole range of voltages as more channels get hit with increasing
gas gain. Above Vd=2750V most of the rate is produced by
coincident hits of several channels.

\begin{figure}[htb]
\centering
\includegraphics[width=0.7\textwidth]{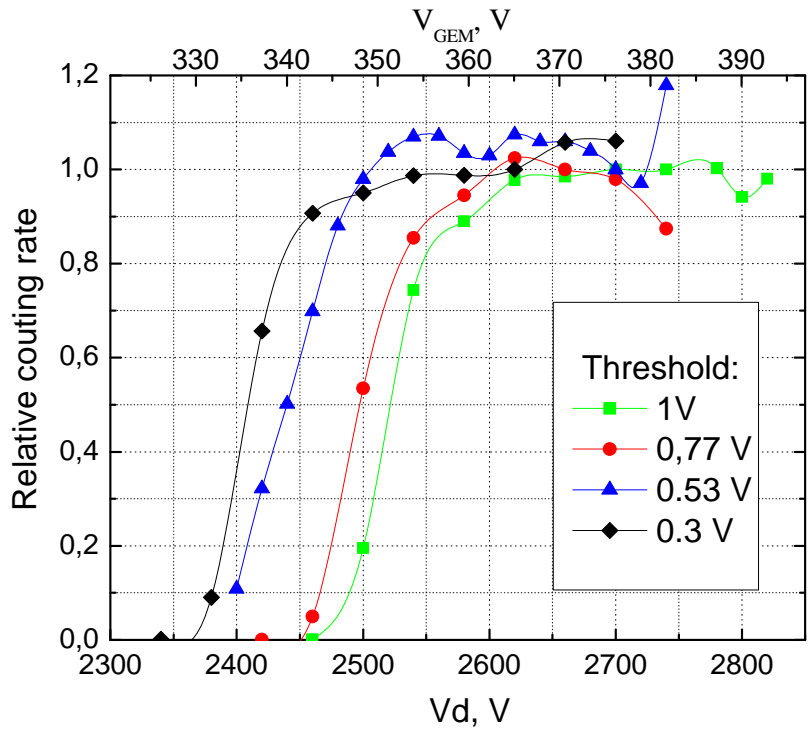}
\caption{Counting rate as a function of voltage at the resistive
divider while the detector is irradiated by thin beam of 8.3keV
photons. The voltage across each GEM is shown at the top scale.
Several dependences corresponding to different comparator
thresholds are shown. } \label{fig:SR-count}
\end{figure}

However the counting rate dependence on voltage looks differently
if the detector is irradiated with narrow X-ray beam. If all the
photons are absorbed within one channel and no charge can come
from the neighboring areas, the counting rate becomes constant
when all the signals induced by absorbed photons become larger
than the threshold. In Fig.~\ref{fig:SR-count} the counting rate
vs voltage dependence for several comparator thresholds are
presented. The 20$\mu$m beam of 8.3 keV photons hits the center of
the channel in this measurement. The starting points of counting
rate plateau give information about average signal value at a
given voltage, thus the gain-voltage characteristic can be derived
from this data. Table ~\ref{tab:Plato} summarizes data on counting
rate plateau starting voltages (50$\%$ level of plateau),
corresponding threshold values and gain values, calculated using
electronic calibration.

\begin{table}[htb]
\centering
\begin{tabular}{|lccc|}
\hline
Plateau starting voltage, V & Threshold, V  & Gain &\\
\hline
 2420  & 0.3 & 450   &\\
 2480  & 0.53& 795   &\\
 2500  & 0.77& 1155  &\\
 2520  & 1.0 & 1500  &\\
\hline
\end{tabular}
\caption{Counting rate plateau starting voltage (50$\%$ level),
corresponding comparator threshold and gain value, calculated from
electronic calibration.} \label{tab:Plato}
\end{table}

\begin{figure}[htb]
\centering
\includegraphics[width=0.7\textwidth]{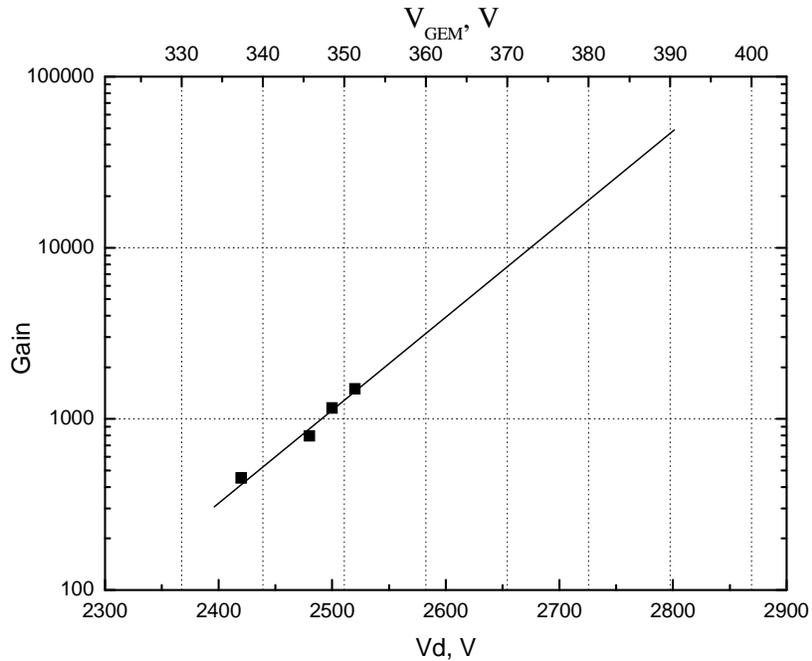}
\caption{Gain as a function of voltage at the divider (bottom
scale) and single GEM(top scale). Exponential fit is plotted
through the experimental points. \newline \newline }
\label{fig:OD4-gain}
\end{figure}

This data is plotted in Fig.~\ref{fig:OD4-gain} with exponential
fit through the experimental points. The gain-voltage dependence
fits well to similar data from ~\cite{GEM-highgain} and
~\cite{OD4-paper1}.

\begin{figure}[htb]
\centering
\includegraphics[width=0.7\textwidth]{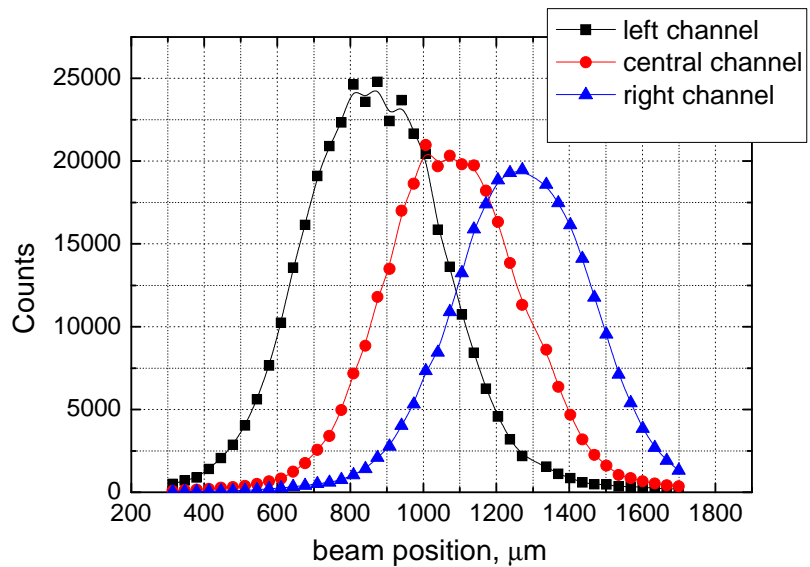}
\caption{Channel response curves for 3 channels scanned with
20$\mu$m wide X-ray beam. Voltage at the divider Vd=2580V, the
comparator threshold provides 90$\%$ efficiency in the central
channel. } \label{fig:Channels}
\end{figure}

For the studies of spatial resolution the detector has been moved
horizontally in such a way that 20$\mu$m wide X-ray beam has
scanned the area of several detector channels. Counting rate as a
function of beam position for each channel (channel response
curve) has been used for characterization of the spatial
resolution. The measurements have been performed at a different
detector voltages and different comparator thresholds in order to
observe the dependence of resolution on these parameters. An
example of the set of channel response curves for Vd=2580V
(V$_{GEM}\sim$360V, Gain$\sim$3000) and the threshold that
provides 90$\%$ efficiency in the central channel is shown in
Fig.~\ref{fig:Channels}.

\newpage

\begin{figure}[htb]
\centering
\includegraphics[width=0.7\textwidth]{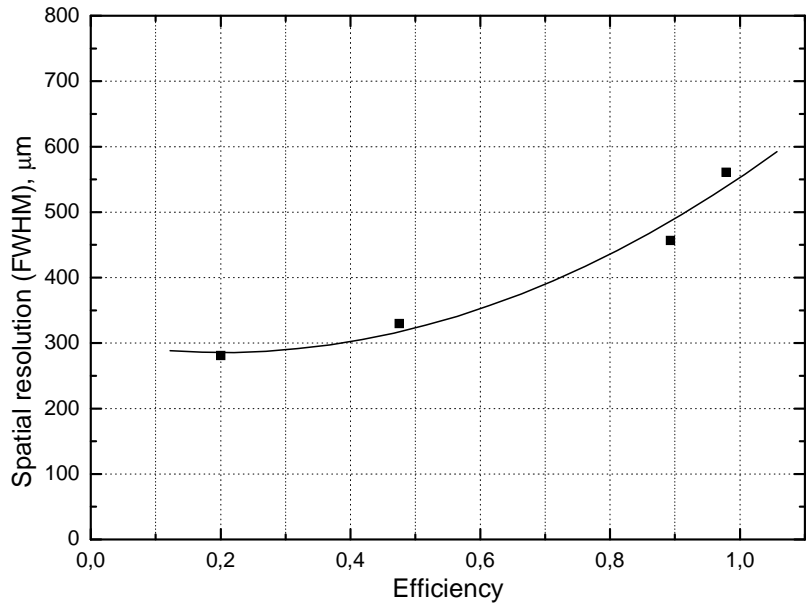}
\caption{Spatial resolution (FWHM of channel response curve) as a
function of efficiency. Voltage at the divider Vd=2580V. \newline
\newline \newline } \label{fig:Resolution-Efficiency}
\end{figure}

Different counting rate of the left and central channels can be
explained by different gain in these areas. The left channel has
slightly higher gain and thus reach full efficiency already at
this voltage and threshold. The lower is the threshold the smaller
signal can be detected and thus the channel response curve is
becoming wider. On the other hand when the threshold is too high
the efficiency starts to drop.
Fig.~\ref{fig:Resolution-Efficiency} demonstrates the dependence
of spatial resolution (FWHM of channel response curve) on the
efficiency, derived from several measurements of channel response
curves. All these measurements have been done at Vd=2580V.

Thus spatial resolution can be tuned in a wide range by the
adjustment of the comparator threshold and/or the detector gain.
At the level of 90$\%$ efficiency the resolution is close to
470$\mu$m and can be improved to FWHM$\sim$330$\mu$m at the
expense of the efficiency that drops down to 50$\%$ in the latter
case.

\begin{figure}[htb]
\centering
\includegraphics[width=0.7\textwidth]{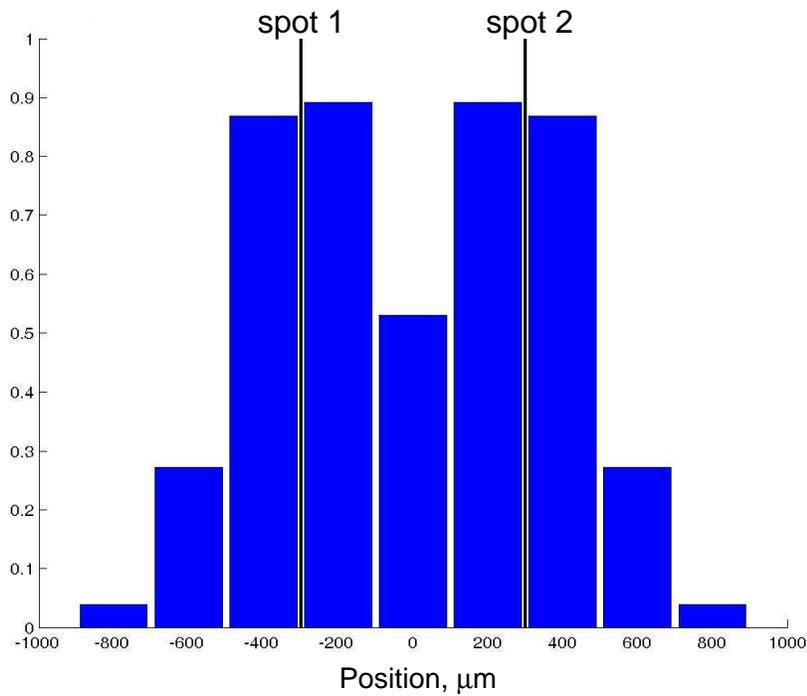}
\caption{Image of two diffraction spots separated by 0.1 degrees
(0.6mm at 350mm distance to the scattering source). Image is
calculated using channel response curve with 470$\mu$m FWHM.}
\label{fig:Diffraction}
\end{figure}

Spatial resolution obtained with 90$\%$ efficiency is good enough
to allow separation of two diffraction spots at angular distance
of 0.1 degree that corresponds to $\sim$0.6mm for this detector.
The image of two such spots positioned symmetrically with respect
to the central channel is calculated using the channel response
curve from Fig.~\ref{fig:Channels} and shown in
Fig.~\ref{fig:Diffraction}.

High rate capability is one of the advantages of GEM based
detectors over wire chambers. Smaller amplifier cells (140$\mu$m
distance between holes in GEM) allow faster charge removal and
thus produce lower space charge that affects the gain. Rate
capability of cascaded GEMs was demonstrated up to the level of
10$^5$ Hz/mm$^2$ of 8 keV photons (~\cite{GEM-rate}). In OD4 we
have been aiming to get the counting rate capability of up to
100kHz per 0.2mm wide channel.

The measurement of rate capability of OD4 was performed with
20$\mu$m wide beam of 8.3keV photons aligned at the center of a
channel. The beam was attenuated with 50$\mu$m thick aluminum
foils. In each subsequent measurement 1 foil was removed to
increase the rate. Every time when the measurement with reduced
number of foils was completed the additional normalizing
measurement was performed with 12 foils. The normalizing
measurement was necessary to correct the results for beam
intensity variations that happened due to monochromator movements
and electron beam instabilities. All the measurements have been
done at Vd=2580V (gain$\sim$3000) and the comparator threshold
adjusted to provide 90$\%$ efficiency of plateau level.

After the completion of all the measurements and correction on the
normalizing data, the effective absorption of the foil has been
calculated for each subsequent pair of measurements. This value
has included the absorption by itself and possible additional rate
reduction due to limited rate capability. Then the average
effective foil absorption and its variance have been calculated
using only the set of measurements where foil absorptions have
been constant (at lower rates). The linear rate scale has been
calculated using the rate value in the first measurement at the
lowest rate and average foil absorption. The relative efficiency
has been obtained as the ratio of the measured rate and the
calculated linear rate. The variance of foil absorption value has
been used to calculate the variance of the linear rate and the
corresponding variance of the relative efficiency.

\begin{figure}[htb]
\centering
\includegraphics[width=0.7\textwidth]{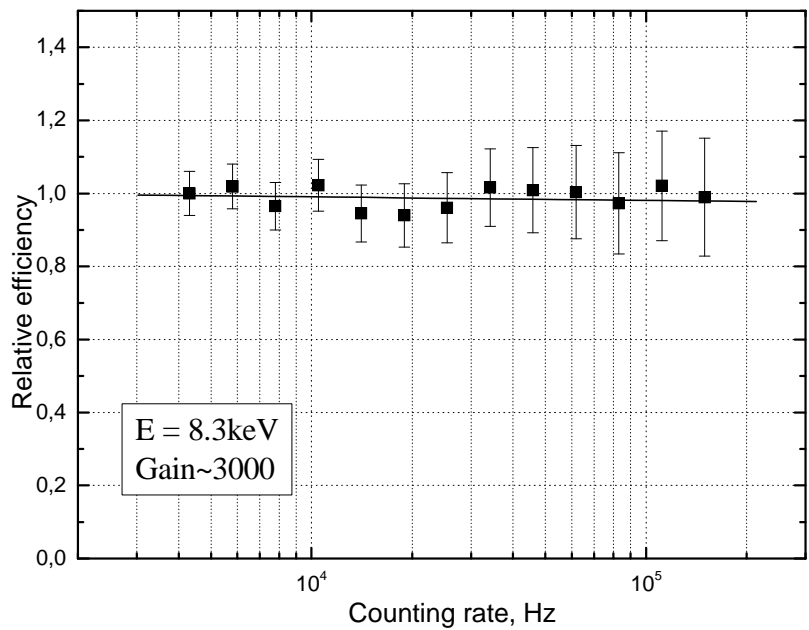}
\caption{Rate capability of OD4. The measurement has been
performed with 20$\mu$m wide X-ray beam aligned at the center of a
channel.} \label{fig:Rate}
\end{figure}

 The result of this study is shown in Fig.~\ref{fig:Rate}.
Within the measurement errors that have been mainly determined
with the beam instabilities, the detector efficiency does not
depend on the photons rate up to $\sim$100kHz/channel. Indeed we
could expect this result because the effective area where charge
is produced by the X-ray beam is spread over 0.2mm*30mm=6mm$^2$
(0.2mm is channel width, 30mm is strip length) and maximum charge
rate is equivalent to only $\sim$20kHz/mm$^2$ of 8.3keV photons at
gain$\sim$3000.

\section{Conclusion.}

Full size detector for WAXS studies has been assembled and tested
at the synchrotron radiation line at VEPP-3. The detector has been
partially equipped with electronics that included
amplifier-shaper, comparator and scaler in each channel. OD4 has
demonstrated stable performance with 8.3 keV photons in the range
of gains from $\sim$500 to more than 10000.

Spatial resolution of the detector can be tuned with the
comparator threshold and gas gain and it appeared to be simple
function of efficiency when the latter is lower than 100$\%$. For
90$\%$ efficiency the spatial resolution is $\sim$470$\mu$m (FWHM
of the channel response curve). Such resolution is enough to
separate clearly the diffraction spots at an angular distance of
0.1 degree, that was initially established as a main requirement
for OD4. The measurements of spatial resolution and efficiency in
a wide range of gas gains and comparator thresholds have shown
that the detector can work at rather low gains (below 1000) and by
threshold adjustment the resolution and efficiency can be chosen
at optimal level. We hope that for the operation at such gains the
number of GEMs in the cascade can be reduced to 2 or even to 1 and
this will be checked in future studies.

Rate capability of OD4 was tested up to the rate of
$\sim$150kHz/channel. No significant degradation of efficiency has
been detected.

The electronics that have been used for the measurements was not
final. The boards that contain the amplifier-shapers and
comparators will be connected to the motherboard that will collect
data from all the 32-channel amplifier boards, control them and
communicate with the computer. The first version of the final
electronics including the motherboard and amplifier cards for
total number of 256 channels will be ready during 2008.

\newpage

\end{document}